\documentclass[a4paper, 12pt, margin=1in]{article}
\usepackage{amsmath,amsthm,amssymb,mathrsfs,graphicx,url}
\usepackage{fullpage}
\usepackage{multirow}
\usepackage[usenames]{color}
\usepackage{ifthen}
\usepackage{setspace}
\usepackage{caption}
\usepackage{subfigure}
\usepackage{float}
\usepackage{graphicx}
\usepackage{algorithm}
\usepackage[noend]{algpseudocode}
\usepackage{times}
\usepackage{bm}
\usepackage{bbm}
\usepackage{natbib}
\usepackage{caption}
\usepackage{float}
\usepackage{subfigure}
\usepackage{eqnarray,amsmath}
\usepackage{color}
\usepackage{multirow}
\usepackage{bbm}
\usepackage[normalem]{ulem}
\usepackage{wrapfig}
\usepackage{authblk}
\usepackage{natbib}

\makeatletter
\def\BState{\State\hskip-\ALG@thistlm}
\makeatother

\theoremstyle{plain}

\theoremstyle{remark}

\theoremstyle{definition}

\theoremstyle{remark}

\theoremstyle{definition}

\renewcommand{\phi}{\varphi}

\title{Modeling Brain Connectivity with Graphical Models on Frequency Domain}
\author[1]{Xu Gao\thanks{xgao2@uci.edu}}
\affil[1]{Department of Statistics, University of California, Irvine, CA, USA}
\author[1]{Weining Shen}
\author[2,5]{Chee-Ming Ting}
\affil[2]{School of Biomedical Engineering \& Health Sciences, Universiti Teknologi Malaysia, Malaysia}
\author[3]{Steven C. Cramer}
\affil[3]{Department of Anatomy and Neurobiology, University of California, Irvine, CA, USA}
\author[4]{Ramesh Srinivasan}
\affil[4]{Department of Cognitive Sciences, University of California, Irvine, CA, USA}
\author[5]{Hernando Ombao}
\affil[5]{Statistics Program, King Abdullah University of Science and Technology, Thuwal, Saudi Arabia}

\begin{document}

\maketitle

\begin{abstract}
Multichannel electroencephalograms (EEGs) have been widely used to study 
cortical connectivity during acquisition of motor skills. In this paper, we introduce copula Gaussian graphical models on spectral domain to characterize 
dependence in oscillatory activity between channels. To obtain a simple and robust representation of brain connectivity that can explain the most variation in the observed signals, we propose a framework based on maximizing penalized likelihood with Lasso regularization to search for the sparse precision matrix. % that \footnote{To XU: we need to explain why we want the sparse precision matrix ... do we want the simplest representation of brain connectivity that can explain the most variation in the 
%observed signals?} obtained from 
To address the optimization problem, graphical Lasso, Ledoit-Wolf and sparse estimation of a covariance matrix (SPCOV) algorithms were modified and implemented. Simulations show the benefit of using the proposed algorithms in terms of robustness and small estimation errors. %\footnote{XU: please elaborate on this "small error" ... what is "error" here? Prediction error (misclassification) or estimation error.}. 
Furthermore, analysis of the EEG 
data in a motor skill task %\footnote{XU: please be explicity ... EEG data in a motor skill task???}, 
conducted using algorithms of modified graphical LASSO and Ledoit-Wolf, 
reveal a sparse pattern of brain connectivity among cortices
which is consistent with the results from other work in the literature.%\footnote{To XU: here I read leading phrases like "we will implement" ... "we will use" ... a referee could catch these and wonder what exactly was the novelty... we need an emphatic statement that reads like "Our main contribution here is ..."}
\end{abstract}


\section{Introduction}
A graph is a model representation of a complex system determined by a set 
of nodes (vertices) and edges connecting them \citep{Trudeau:1993}. On the foundation of graph and probability theory, graphical models (probabilistic graphical model) are widely used in the communities of Bayesian statistics, 
statistical learning and machine learning. In the framework of graphical models, each node represents a random variable and edges denote the probabilistic relationship between nodes. The graph depicts the structure  where the joint distribution of random variables can be decomposed into a product of factors depending only on subsets of variables \citep{Bishop:2006}.

Graphs have been introduced to model brain connectivity through the way that nodes represent cortical and subcortical regions while edges can be denoted as functional and structural connections between cortical nodes \citep{Bullmore:2011}. In the literature of brain graph models, much work has been done for all major modalities of functional magnetic resonance imaging (fMRI) and neurophysiological data. To name a few, functional brain graphs and their relevant works have been constructed from fMRI \citep{Achard:2007}, electroencephalography (EEG) signals \citep{Michelovannis:2006, Kemmer:2015}, magnetoencephalography (MEG) data \citep{Stam:2004} and Local Field Potentials (LFPs) signals \citep{Gao:2016, Gao:2018}. From diffusion tensor imaging (DTI) and diffusion spectrum imaging(DSI), structural brain graphs have been studied by \citet{Gong:2008}. Among them, sparse graphical models, which are widely discussed in \citet{Friedman:2008}, are highly efficient in inferring multielectrode brain recordings. The sparsity of graph provides a robust approach that highlights the most significant interactions between brain cortices and helps to interpret the data \citep{Dauwels:2012}.%\footnote{To XU: you can lift some phrases from the last sentence as the motivation for sparsity and use that in the 
%abstract.}

Motivated by the advantages of using sparse graph, we considered the problem of modeling brain connectivity from EEG data under the framework of sparse graphical models. Inspired by the work of \citet{Dauwelsyu:2012}, the main contributions of the paper are as follows: (1.) We introduce copula Gaussian graphical models to account for the non-Gaussianity of signals on frequency domain; (2.) We develop a framework to capture the between-channel dependence in the oscillatory activity; (3.) By including a regularization term, we are able to obtain a simple and robust representation that uncovers the most critical connectivities between brain cortices. 
%Due to the dependence property of EEG data, we follow the framework of graphical models in multivariate time series on frequency domain \citep{Dauwelsyu:2012}.
%\footnote{To XU: emphasize the novelty ... is there anything about developing graphical models to capture between-channel dependence in the oscillatory activity??? Also need to de-emphasize Dauwelsyu ... just say that this work is inspired by that paper but mention the extensions, differences, etc.}

In this paper, the core methodology boils down to the sparse estimation of covariance matrices. Specifically, we regularize the log likelihood with a lasso penalty on the entries of covariance matrix as the objective function. The penalty %\footnote{To XU: there is a philosophical issue here. To me the true underlying connectivity is NOT sparse. Everything is connected to everything. However many of these connections are weak. Thus the goal of any modeling that imposes sparsity is that it finds those strongest and most meaningful connections. To be clear though: sparsity methods should not claim that the true underlying connectivity structure is sparse.} 
is used to reduce the effective number of brain connectivity and thus produces sparse and robust estimates \citep{Bein:2010}. To solve the optimization problem, many algorithms have been introduced in the literature. \citet{Rothman:2008} propose a novel algorithm with the assumption of ordering to the variables. \citet{Butte:2000} introduce relevance work in working on the optimization problem in the way that pairwise correlation beyond a threshold are linked by an edge. \citet{Rothman:2009} propose an algorithm by introducing shrinkage operators. \citet{Rothman:2010} utilize lasso-regression based method to solve the optimization problem. Throughout this paper, we develop modified algorithms based on the works of graphical lasso \citep{Friedman:2008}, sparse estimation of covariance proposed by \cite{Bein:2010} and LedoitWolf algorithm \citep{Ledoit:2012}.

The paper is organized as follows. In section \ref{background}, we briefly review copula Gaussian graphical models and its application in EEG data. In section \ref{statement}, we formulate the optimization problem and propose three algorithms to solve it. In section \ref{simu}, we conduct simulations to test the robustness and performance of the three algorithms. In section \ref{real}, we apply the EEG data to the optimization problem in searching for the sparse connectivity matrix.
\section{Background on Graphical Models in Brain Connectivity}
\label{background}
In this section, we discuss the preliminaries on graphical models and its application in modeling brain connectivity from EEG data.
\subsection{Copula Gaussian Graphical Model}
Suppose we have non-Gaussian random variables $Y_1, Y_2, \cdots, Y_n$. We define hidden Gaussian random variables $X_1, X_2, \cdots, X_n$ through the relationship that \citep{Dauwels:2012}
\begin{align}
X_k \sim \mathcal{N}(0, \Sigma_k),\\
Y_k = F_k^{-1}(\Phi(X_k)),
\end{align}
where ${\Sigma_k^{-1}}$ is the precision matrix, $\Phi$ is the cumulative distribution function of Gaussian random variables and $F_k$ is the empirical cumulative distribution function of $Y_k$. In practice, $F_k^{-1}$ can be estimated by
$$
F_k^{-1}(y)=\inf\{z, F_k(z) \geq y\}.
$$
\subsection{Graphical Models in EEG data}
In fact, the EEG data obtained from different electrodes is multiple non-Gaussian time series. By implementing Copula Gaussian Graphical Models, the original time series have been transformed into Gaussian dataset. In the following work, we implement methodology in graphical models for multivariate Gaussian time series.

Suppose a graphical model $G=(V, E)$ uniquely determines the conditional independence on Gaussian process $X(t)=(X_1(t), \cdots, X_p(t)).$ In graph $G$, each node $V_i$ denotes a single time series $X_i(t).$ The absence of edge between $V_i$ and $V_j$ denotes the conditional independence between time series $X_i(t)$ and $X_j(t)$. Under the assumption that the cross-variance function of $X(t)$ is summable, $$\sum \limits_{\tau=-\infty}^\infty |cov\{X_i(t), X_j(t+\tau)\}|<\infty, \forall i, j,$$ we define the cross spectral density matrix of $X$ as
$$
S_{i,j}(\omega)=\mathcal{F}\{cov(X_i, X_j)\},
$$
where $\mathcal{F}$ denotes the Fourier transform. As a result of \citet{Dahlhaus:2000}, the Gaussian process $X_i$ and $X_j$ are conditional independence if and only if
$$
\{S(\omega)^{-1}\}_{ij}=0, \quad \forall \omega.
$$
%We may assume that the cross spectral density matrix $S(\omega)^{-1}$ plays a similar role the precision matrix in graph $G.$ 
In practice, we use $S(\omega)$ as the empirical variance-covariance matrix of the time series $X(t).$
\section{Optimization Problem Statement and the Proposed Algorithms}
\label{statement}
Following the methodology discussed in section \ref{background}, we have transformed the EEG data into quasi-Gaussian time series with empirical variance-covariance matrix $S(\omega).$ In this section, we formulate the final optimization problem in modeling the brain connectivity and propose three algorithms in solving the problem.
\subsection{Optimization Problem Formulation}
In sparse graphical models, true brain connectivity involving the strongest and the most relevant connections is uniquely determined by the sparse precision matrix (the inverse of covariance matrix). %Under the framework of \citet{Dauwels:2012}, the objective is the negative log-likelihood function with Lasso penalty on the entries of the precision matrix. In particular, %we assume that the observed data $X=(X_1(t), \cdots, X_p(t)$ follows multivariate Gaussian distribution with mean $\mu$ and covariance matrix $\Sigma$, i.e.
%$$
%X_i(t)\sim MVN(\mu, \Sigma)\quad \forall i,
%$$
The objective function, defined as the regularized negative log-likelihood function is given by
\begin{equation}
\begin{aligned}
& \underset{\Sigma}{\text{minimize}}
& & -\log \det(\Sigma) + \text{tr} (S(\omega)*\Sigma)+\lambda*||\Sigma||_1 \\
& \text{subject to}
& & \Sigma \succeq 0,
\end{aligned}
\label{optim}
\end{equation}
 where $S(\omega)$ is the empirical variance-covariance matrix discussed in section \ref{background}, $||\Sigma||_1$ is the $l_1$ norm which is given by the sum over the absolute values of entries in matrix $\Sigma,$ and $\lambda$ is a tuning parameter controlling the amount of $l_1$ shrinkage.
\subsection{The Proposed Algorithms}%\footnote{To XU: need to emphasize the novelty discussed in this section ... add this to the abstract and intro and conclusion}
We propose three algorithms to address the optimization problem (\ref{optim}).%, we mainly focus on three algorithms.
%\subsubsection{SPCOV (Majorize-Minimize) algorithm}

\underline{SPCOV (Majorize-Minimize) algorithm}

In the objective function (\ref{optim}), $\text{tr} (S(\omega)*\Sigma)+\lambda*||\Sigma||_1$  is convex while $ \log \det(\Sigma)$ is concave, thus a majorize-minimize scheme could be used. In summary, this algorithm presents as two loops, the outer loop
approximates the non-convex problem and the inner loop solves each convex
relaxation.

\underline{Graphical lasso algorithm}

We also propose a modified graphical lasso algorithm. The rationale is that suppose $\hat{\Sigma}$ is the estimate of  $\Sigma$, then one can solve the problem by optimizing over each row and corresponding column of $W$ in a block coordinate descent fashion. 

\underline{Modified Ledoit-Wolf algorithm}

Inspired by the works of \cite{Ledoit:2012}, \cite{Fiecas:2010} and \cite{Fiecas:2011}, we also implement a modified Ledoit-Wolf algorithm to address the optimization problem \ref{optim}. Specifically, we utilize the sample covariance $S$ and the maximum likelihood estimator $S^{ML}$ to obtain a shrinkage estimator which compromises between variance and bias. The shrinkage intensity is evaluated by minimizing a risk function based on mean square errors.

\section{Simulations}
\label{simu}
In this section, two simulation scenarios were considered to evaluate the performance of the proposed algorithms. In each scenario, we created different types of sparse symmetric positive definite matrices to be the true covariance structure. We randomly generated 200 samples from the true covariance structure and then used the data as the input of the proposed algorithms. Each scenarios was repeated 1000 times. We evaluated the results by its ability to correctly identify the zero elements of $\Sigma$ and the discrepancy with the true covariance matrix by root- mean-square error, $||\hat{\Sigma}-\Sigma||_{F}/p$ and entropy loss, $-\log\det(\hat{\Sigma}\Sigma^{-1})+tr(\hat{\Sigma}\Sigma^{-1})-p$ respectively, where $p$ is the number of parameters.

The first scenario was from Cliques Model. In this model, we set $\Sigma = \text{diag}(\Sigma_1, \Sigma_2, \Sigma_3)$ 
to be the covariance matrix. Each $\Sigma_i$ in the diagonal represented a $6*6$ dense matrix. Other parts of the matrix were zero.
The second scenario was Random Model. The sparse covariance graph was created by assigning $\Sigma_{ij}=\Sigma_{ji}$ to be non-zero with probability $0.02$, independently of other elements.

\begin{table}[h]
\caption{Summary of simulation results for optimization problem (\ref{optim}).  Sample size is chosen as $200$. $1000$ simulations were generated under Random Model and Cliques Modes. For each scenario, we present the empirical average and standard error of Root-Mean-Square Error and Entropy Loss.}
\hskip-1.6cm
\begin{small}
\begin{tabular}{llccccccc}
\hline
&& \multicolumn{3}{c}{Random Model} && \multicolumn{3}{c}{Cliques Model}\\
\cline{3-5} \cline{7-9}
\multicolumn{2}{c}{Method} & \multicolumn{2}{c}{Root-Mean-Square Error} & Entropy Loss  &&  \multicolumn{2}{c}{Root-Mean-Square Error} &  Entropy Loss  \\
\hline

      &Graphical Lasso 	&     \multicolumn{2}{c}{$2.723*10^{-5} \pm 1.020*10^{-6}$}&$0.232 \pm 0.003$  	&&    \multicolumn{2}{c}{$2.730*10^{-5} \pm 1.060*10^{-6}$} &  $0.230 \pm 0.003$ \\
      & Ledoit-Wolf    &    \multicolumn{2}{c}{$5.170*10^{-5}  \pm 6.300*10^{-6}$} &  $0.180 \pm 0.001$   	&&    \multicolumn{2}{c}{$5.270*10^{-5} \pm 6.300*10^{-6}$} 	&   $0.180 \pm 0.001$ \\
     & SPCOV    &    \multicolumn{2}{c}{$7.423 *10^{-5} \pm 9.100 *10^{-6}$} &  $0.732 \pm 0.089$   	&&    \multicolumn{2}{c}{$8.200*10^{-5} \pm 1.370 *10^{-6}$} 	 &   $ 0.700 \pm 0.092$ \\

      \hline
\end{tabular}
\end{small}
\label{table1}
\end{table}

\begin{table}[h]
\caption{Summary of simulation execution time for optimization problem (\ref{optim}).  Sample size is chosen as $200$. $1000$ simulations were generated under Random Model and Cliques Modes. For each scenario, we present the empirical average and standard error of the execution time.}
\centering
\begin{small}
\begin{tabular}{llccccccc}
\hline
&& \multicolumn{3}{c}{Random Model} && \multicolumn{3}{c}{Cliques Model}\\
\cline{3-5} \cline{7-9}
\multicolumn{2}{c}{Method} & \multicolumn{3}{c}{Execution Time}  &&  \multicolumn{3}{c}{Execution Time}  \\
\hline

      &Graphical Lasso 	&     \multicolumn{3}{c}{$0.400 \pm 0.006$}	&&    \multicolumn{3}{c}{$0.330 \pm 0.004$} \\
      & Ledoit-Wolf    &     \multicolumn{3}{c}{$0.007 \pm 5.23*10^{-5}$}	&&    \multicolumn{3}{c}{$0.001 \pm 1.92*10^{-5}$} \\
     & SPCOV    &     \multicolumn{3}{c}{$0.152 \pm 0.033$}	&&    \multicolumn{3}{c}{$0.630\pm 0.227$} \\

      \hline
\end{tabular}
\end{small}
\label{table2}
\end{table}

\begin{figure}[h!] \centering
	\begin{tabular}{c}
		\includegraphics[width=.5\textwidth]{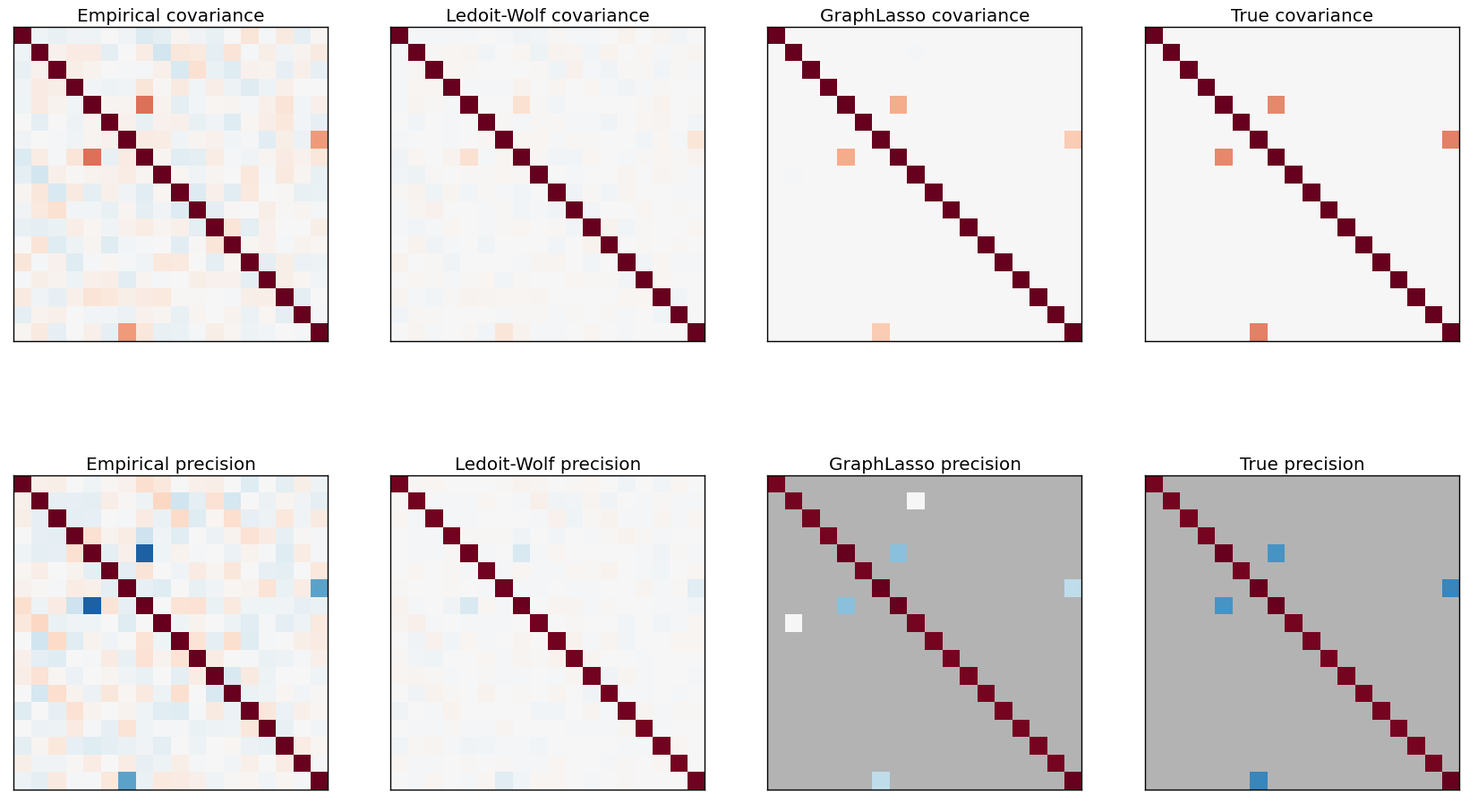} \\
		The connectivity matrix of Random Model \\
			\includegraphics[width=.5\textwidth]{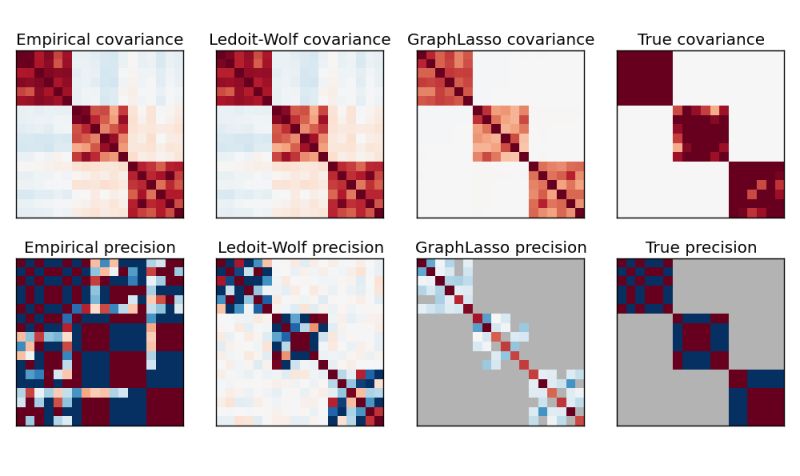} \\
				The connectivity matrix of Cliques Model
	\end{tabular}
	\caption{Visualization of simulation results in Random Model}
	\label{rm1}
\end{figure}

From the Tables \ref{table1} and \ref{table2}, we can see that both Graphic-Lasso and Ledoit-Wolf optimization model had smaller mean square error and entropy loss compared to SPCOV in both scenarios. In addition, all the three algorithms were more accurate and robust in Random Model. In Figure \ref{rm1}, the different shades of color in the connectivity matrix indicated the correlation between each nodes and darker colors demonstrated stronger correlations. It can be seen that the estimated connectivity matrix obtained from Graphical Lasso and Ledoit-Wolf algorithms were in high accordance with the true covariance structure. In summary, simulations suggest that both Graphical Lasso and Ledoit-Wolf methods are competitive for identifying the sparsity structure of the simulated data.

\section{Real-data analysis}
\label{realdata}
\subsection{Dataset description}
\begin{figure}[h!]
\centering
     \includegraphics[width=0.48\textwidth]{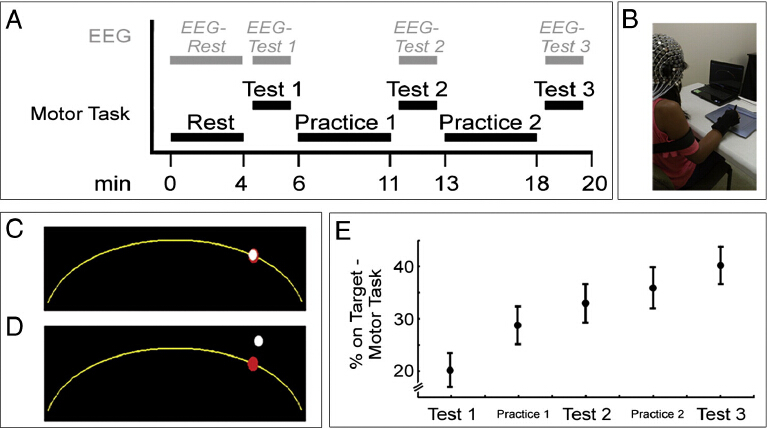}
     \caption{Experiment setup}\label{ex}
\end{figure}

\begin{figure}[h!]
    \centering
     \includegraphics[width=10cm,height=4.5cm]{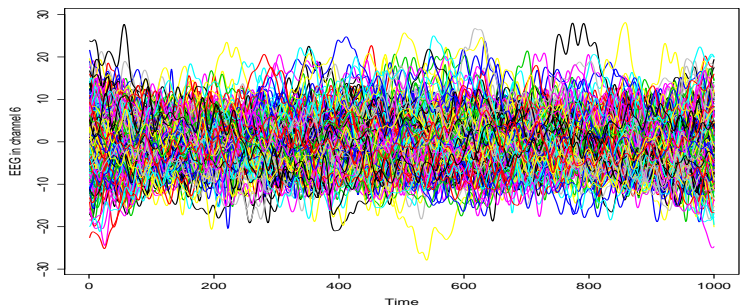}
     \caption{EEG signals across trials in channel 6}\label{eeg}
\end{figure}

The EEG dataset was obtained from a multi-subject stroke experiment conducted at the University of California Irvine Neurorehabilitation Lab (PI: Cramer). %\footnote{To XU: (1.) is this from Jennifer Wu's dataset? If yes we need to cite that paper and also the lab by Steve C. Cramer. (2.) should we include Steve Cramer as a co-author? We've done this from Zhe Yu's papers. } 
During the experiment, participants sat in a chair facing a monitor in a single session. Their task was to make movements across different centers of each circle on the screen. To minimize the variability between individuals, the researchers measured the awake resting-state EEG for 3 min (EEG-Rest) at 1000 Hz prior to the motion task. Then, the measurement of each participant's maximum arm movement speed was obtained, and a baseline assessment of motor skill task was recorded. During this procedure, EEG was measured (EEG-Test1). Later on, the participant was required to receive a practice block, followed by another test block. Finally, after three tests and two practice blocks were done, the EEG was obtained, which comprised of four scenarios -- EEG-Test(1-3) and EEG-Rest. Figure \ref{ex} shows the general setup of this experiment.
There are 16 subjects recorded in the entire dataset. As for each subject, each scenario of dataset consists of 160 trials, 1000 time points and 256 channels indicating different cortical regions. In this paper, subject ``YUGR'' is chosen. Figure \ref{eeg} shows the EEG signal across trials in channel 6 from subject ``YUGR''. The dataset consists of 160 epochs from EEG-Rest, 73, 74 and 63 trials from EEG-Test 1-3 respectively.

\subsection{Modeling and results}
To study the brain connectivity during the motion experiment, we applied the EEG data to the optimization problem (\ref{optim}). Motivated by the results from simulations, Graphical Lasso and Ledoit-Wolf algorithms were implemented to search for the solution. Figure \ref{real} shows the connectivity matrix across different brain cortices from the two algorithms. It can be found that both of the matrices show a similar pattern in regrading to the correlation between cortices. In particular, the total cortices can be classified as four regions in which high association can be realized.%\footnote{To XU: compare and contrast these results from J Wu's paper. }

 \begin{figure}[h!] \centering
	\includegraphics[width=.5\textwidth]{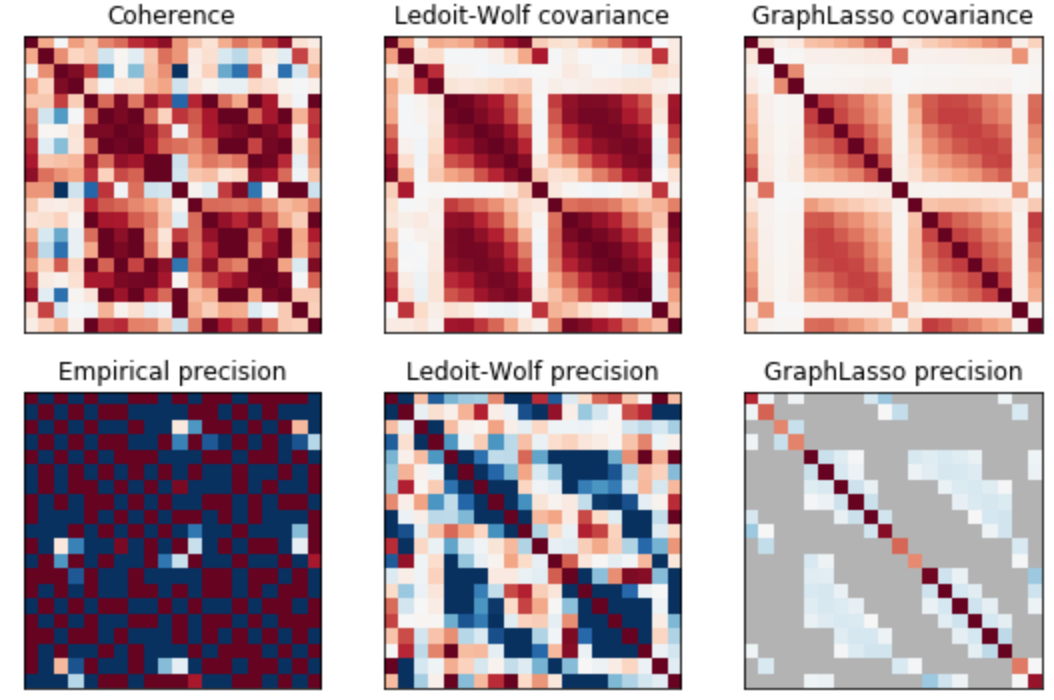} 
	\caption{	The connectivity matrix of the EEG data }
	\label{real}
\end{figure}

\iffalse
 \begin{figure}[h!] \centering
          \begin{tabular}{cc}
          \includegraphics[width=.5\textwidth]{real1} &
          \includegraphics[width=.5\textwidth]{real2} \\
         The connectivity matrix of the EEG data & Model selection plot
          \end{tabular}
          \caption{Visualization of the results in EEG data}
            \label{real}
          \end{figure}
\fi

\section{Conclusion}

We have developed a method to model brain connectivity through graphical models based on the framework of Copula Gaussian Model. To further address the optimization problem (\ref{optim}), Graphical Lasso, Ledoit-Wolf and SPCOV algorithms were implemented and compared. Simulations results show the benefit of using Graphical Lasso and Ledoit-Wolf in searching for the connectivity matrix. The proposed algorithms were also conducted on the real EEG data from the motion experiment. Results show the sparsity of the brain connectivity between cortices, which coincides with the results from previous literature.

\end{document}